The enhancement of phase separation aspect in electron doped manganite $Ca_{0.8}Sm_{0.16}Nd_{0.04}MnO_3$

D.A. Filippov[1], K.V. Klimov[1], R.Z. Levitin[1], A.N. Vasil'ev[1], T.N. Voloshok[1], and R. Suryanarayanan[2*]

[1]Physics Faculty, Moscow State University, 119992, Moscow, Russia

[2]Laboratoire de Physico-Chimie de l'Etat Solide, Bât 414 Université Paris-Sud, 91405 Orsay, France

Abstract

The complex lanthanide doping of electron manganites results in enhancement of various phase separation effects in physical properties of these compounds. Selecting $Ca_{0.8}Sm_{0.16}Nd_{0.04}MnO_3$ as a model case we show that the first order structural phase transition from paramagnetic semi-metallic phase into anti-ferromagnetic semi-metallic phase at $T_S \sim 158 \pm 4$ K is marked by an abrupt decrease in magnetization, a step like anomaly $\Delta L/L = 10^{-4}$ in thermal expansion and large latent heat $\Delta Q = 610$ J/mol. In a certain temperature range below $T_S$, the high field magnetization exhibits hysteretic metamagnetic behavior due to field-induced first order transformation. The temperature dependencies of ac-susceptibility, magnetization and resistivity suggest rather a non-uniform state in $Ca_{0.8}Sm_{0.16}Nd_{0.04}MnO_3$ at low temperatures. The metal - insulator transition occurs at $T_{MI} \sim 112 \pm 3$ K, accompanied by the step-like increase in magnetization. These features could be ascribed to "sponging" of electrons from neighboring anti-ferromagnetic matrix by clusters undergoing the ferromagnetic ordering. The spontaneous magnetization of $Ca_{0.8}Sm_{0.16}Nd_{0.04}MnO_3$ at low temperatures is about 0.1% of calculated saturation magnetization for this compound.

PACS numbers: 64.75.+g; 73.43.Qt; 75.47.Lx

*corresponding author: e-mail: ramanathan.suryan@lpces.u-psud.fr

# I. INTRODUCTION

Some of the several factors of primary importance influencing the physical properties of the well known colossal magnetoresistance manganite perovskites $A_{1-x}D_xMnO_3$, are the ratio of Mn 3+ /Mn 4+ ions, the sizes of divalent(D) to trivalent cations(A) substituted at the A site of perovskite structure and the spread in sizes of these cations[1,2]. The ratio of trivalent and tetravalent manganese ions defines the carrier density. This quantity is of key importance in the determination of competition of insulating anti-ferromagnetic and metallic ferromagnetic states. The lengths and angles of Mn - O - Mn bonds in perovskite structure are determined by the size of A cations, while the variance in their sizes is related to disorder effects in these systems [3]. The disorder effects have lead to the understanding that the manganite perovskites are intrinsically non-uniform systems [4,5]. There are multiple evidences that the carrier density is not homogeneously distributed in the patterns of competing and coexisting phases. One of the controversial issues in this respect is the scale of these inhomogeneities. While it is widely accepted [6] that non-uniformity is present at nanoscale level, there are indications on phase separation at the micrometer scale [7].

The compositions with Mn 3+ /Mn 4+ < 1 are called the electron doped manganites. One end member of this family is the compound CaMnO3, which nominally does not possess $Mn^{3+}$ ions but only $Mn^{4+}$. At low temperatures, this compound is an insulating anti-ferromagnet of G-type. Doping it with trivalent lanthanides leads to rather intriguing changes in $Ca_{1-x}Ln_xMnO_3$ series [8-12]. At low x values, the substitution of $Ca^{2+}$ for $Ln^{3+}$ is accompanied initially by the rise of ferromagnetic component in magnetization and increasing conductivity. On further increasing the lanthanide content, the competition between ferromagnetic and anti-ferromagnetic interactions becomes stronger, resulting in peculiar magnetic behavior of electron manganites. In compositions of the most pronounced colossal magnetoresistance effect, to be referred hereafter as optimal, namely at x =0.13 to 0.17, the magnetization exhibits a sharp peak-like anomaly [9]. The peak temperature increases with lanthanides content and decreases with increasing $Ln^{3+}$ ions size. The mixing of different lanthanides substituted into perovskite structure shifts somewhat the optimal x values and broadens the singularity in magnetization. At still higher $Ln^{3+}$ content the heavily electron-doped manganites reach anti-ferromagnetic insulating state of C-type at low temperatures [12].

The origin of weak ferromagnetism in electron doped manganites was attributed to either canting of predominantly anti-ferromagnetic sublattices or to the presence of ferromagnetic clusters embedded into anti-ferromagnetic matrix [11-14]. The neutron diffraction studies seem to support the phase separation scenario [15].The same scenario was used recently to analyse the metastable states in $Ca_{0.85}Sm_{0.15}MnO_3$ [16]. The x = 0.15 composition is optimal in $Ca_{1-x}Sm_xMnO_3$ series [7,18]. On lowering the temperature, the low field magnetization of $Ca_{0.85}Sm_{0.15}MnO_3$ exhibits a sharp and narrow peak at $T_S$= 115 K, accompanied by a metal - insulator transition. These features are due to the structural transformation from high-temperature orthorhombic Pnma phase, which is paramagnetic into low-temperature monoclinic $P2_1/m$ phase, which is an anti-ferromagnet of C-type. Strong magnetic field causes a collapse of low temperature phase in $Ca_{0.85}Sm_{0.15}MnO_3$ resulting in a field induced first order transition from the monoclinic anti-ferromagnetic phase to the orthorhombic ferromagnetic phase [16-19]. The $P2_1/m$ phase at low temperatures shows the signatures of phase separation, the ferromagnetic clusters and anti-ferromagnetic clusters of G-type being present in anti-ferromagnetic matrix of C-type [18]. The effects of phase separation seem to diminish at higher $Ln^{3+}$ content [12].

We report here on the effects of creating additional disorder at the A-site by a rare earth substitution in electron doped manganites $Ca_{1-x-y}Sm_xNd_yMnO_3$. The various physical measurements carried out on the $Ca_{0.8}Sm_{0.16}Nd_{0.04}MnO_3$ sample point out to an enhancement of the phase separation aspect.

## II. SAMPLES AND EXPERIMENTAL SET-UP

The pellets of manganites $Ca_{1-x-y}Sm_xNd_yMnO_3$ were obtained from the stoichiometric mixture of $Sm_2O_3$, $Nd_2O_3$, $CaCO_3$ and $MnO_2$ by direct solid state synthesis. The mixture was heated and reground several times by raising successively the sintering temperature from 980 °C to 1300 °C. The final sintering was done at 1400 °C for 36 hours. The resulting product was checked by X-ray diffraction at room temperature showing no measurable traces of the impurity phases in orthorhombic Pnma matrix. The magnetization $Ca_{1-x-y}Sm_xNd_yMnO_3$ pellets, up to 5 T, was measured as a function of temperature (5 to 300 K) by "Quantum Design" SQUID magnetometer. The magnetization was measured also by an inductive method in pulsed magnetic field up to 20

T. The low-field ($\sim 10^{-3}$ T) ac magnetic susceptibility was measured in a separate set-up at 70 Hz. The resistivity was obtained by a standard four-probe technique and the same set-up was used to measure the thermal expansion by means of resistive strain gauges attached to the sample. The specific heat was measured by quasi adiabatic microcalorimeter.

### III. LOW FIELD MAGNETIZATION AND AC SUSCEPTIBILITY

The temperature dependencies of magnetization of $Ca_{1-x-y}Sm_xNd_yMnO_3$ of four different compositions taken during heating in a magnetic field 0.1 T are shown in Fig. 1. The magnetization of $Ca_{0.85}Sm_{0.08}Nd_{0.07}MnO_3$ and $Ca_{0.85}Sm_{0.10}Nd_{0.05}MnO_3$ samples (in both cases x + y = 0.15) exhibits sharp peaks at $T_S \sim 119$ K ($\sim 122$ K) and subsequent gradual increase at lowering temperature. The magnetization of $Ca_{0.80}Sm_{0.16}Nd_{0.04}MnO_3$ and $Ca_{0.80}Sm_{0.18}Nd_{0.02}MnO_3$ samples (in both cases x + y = 0.20) shows broader peak at $T_S \sim 162$ K ($\sim 171$ K), abrupt decrease of the moment below this temperature, and step-like upturn at lower temperatures. The magnitude of magnetization at peak's temperature in (x + y) = 0.15 samples is somewhat lower than that found (x + y) = 0.20 samples.

The magnetization of $Ca_{0.80}Sm_{0.16}Nd_{0.04}MnO_3$ sample taken at increasing and decreasing temperature possesses the hysteretic features, as shown in Fig. 2. Most clearly these features are seen in the vicinity of peak and at lowest temperatures. There is no appreciable hysteresis at step-like anomaly. Two anomalies of similar shape are seen at $T_{MI} \sim 109$ K and $T_S \sim 162$ K in ac magnetic susceptibility, shown in the Fig. 2(inset). The field dependencies of magnetization in $Ca_{0.80}Sm_{0.16}Nd_{0.04}MnO_3$ sample are linear at high temperatures, as shown in Fig. 3a. In a temperature range between peak-like and step-like anomalies the deviations from linearity (H) curves gradually develop, but weak ferromagnetic component in magnetization appears only in vicinity of step-like upturn (Fig. 3b). The magnetic hysteresis loop taken at T = 5 K is shown in Fig. 3b(inset). The slope of the magnetization curves at high field was used to determine high field magnetic susceptibility shown in Fig. 4. The temperature dependence of magnetic susceptibility in high temperature region follows Curie-Weiss law with a paramagnetic Curie temperature $\Theta \sim 133$ K, indicating a predominance of ferromagnetic interaction in high temperature phase of this compound. As seen from the inset to Fig. 4 the spontaneous magnetization at low temperatures is no more than 0.1% of saturation magnetization $\sim 4$ $\mu_B$ for this compound.

## IV. HIGH FIELD MAGNETIZATION

The measurements of the magnetization of $Ca_{0.80}Sm_{0.16}Nd_{0.04}MnO_3$ in high magnetic fields show that the application of a magnetic field at low temperatures induces a metamagnetic transition. The field dependencies of magnetization of $Ca_{0.80}Sm_{0.16}Nd_{0.04}MnO_3$ taken in pulsed magnetic field are shown in Fig. 5. These dependencies qualitatively differ at crossing the peak's temperature. Above $T_S$, the magnetization is almost linear at low fields and experiences the tendency to paramagnetic saturation at high fields. Below $T_S$, the magnetization is metamagnetic in character and shows the pronounced hysteresis at transition from a low moment state to a high moment state. The critical field of metamagnetic transition increases on lowering the temperature. At low temperatures the metamagnetic transition is not reached in the magnetic field range studied, so that magnetization just gradually rises on application of magnetic field.

## V RESISTIVITY, THERMAL EXPANSION AND SPECIFIC HEAT

The measurements of resistivity, thermal expansion and specific heat were done in the absence of a magnetic field. The resistivity of $Ca_{0.80}Sm_{0.16}Nd_{0.04}MnO_3$ sample, shown in Fig.5(panel a) decreases by four orders of magnitude on heating in the range 80-300 K. There are two distinct chnages of slope or kinks at $T_{MI} \sim 115$ K and $T_S \sim 158$ K, both leading to decrease in resistivity. While the kink at $T_S$, separates two different regions of semi-metallic behavior, that at $T_{MI}$ marks the metal - insulator transition. The thermal expansion of $Ca_{0.80}Sm_{0.16}Nd_{0.04}MnO_3$ sample, shown in Fig. 5 (panel b), is almost linear in a range 80 - 300 K, except the step like anomaly $\Delta L/L \sim 10^{-4}$ at $T_S \sim 158$ K. The temperature dependence of specific heat, shown in Fig. 5 (panel c), evidences well defined peak at $T_S \sim 158$ K while showing no traceable singularities at any other temperature. The area between the experimentally obtained curve and polynomial fitting curve was used to estimate the latent heat of phase transition Q = 610 J/mol.

## VI. DISCUSSION

Based on our experimental data and referring to similar measurements performed on Nd-free $Ca_{1-x}Sm_xMnO_3$ samples[8-10,17-19], the succession of transitions in $Ca_{0.80}Sm_{0.16}Nd_{0.04}MnO_3$ can be understood as follows.

The anomalies of physical properties at $T_S \sim 158 \pm 4$ K in $Ca_{0.80}Sm_{0.16}Nd_{0.04}MnO_3$ are similar to that observed in $Ca_{0.85}Sm_{0.15}MnO_3$ at structural transformation from high temperature orthorhombic Pnma phase to low temperature monoclinic $P2_1/m$ phase. Presumably, a similar transition takes place at complex doping in the present Sm-Nd doped sample. The latent heat of transition in $Ca_{0.80}Sm_{0.16}Nd_{0.04}MnO_3$ is comparable with the latent heat Q = 500 J/mol released[19] during the structural phase transition in $Ca_{0.85}Sm_{0.15}MnO_3$[19]. The application of high magnetic field at low temperatures results in a field induced structural transformation of first order from monoclinic $P2_1/m$ phase to orthorhombic Pnma phase. This is evident from the metamagnetic character of magnetization curves with a large hysteresis. The specific heat calculated from the jump in magnetization at metamagnetic transition is in good correspondence with latent heat released at the spontaneous phase transition.

At $T > T_S$, the sample is paramagnetic and the magnetic susceptibility follows the Curie-Weiss law with paramagnetic Curie temperature $\Theta = 133$ K and effective magnetic moment $\mu_{eff} \sim 4$ $\mu_B$. This value is consistent with the estimate for a given combination of spin only magnetic moments of $Sm^{3+}$, $Nd^{3+}$, $Mn^{3+}$ and $Mn^{4+}$ ions in the sample. The predominance of ferromagnetic exchange interaction in high temperature orthorhombic Pnma phase would lead to the formation of ferromagnetic state in this phase, if the structural transition to anti-ferromagnetic monoclinic $P2_1/m$ phase had not occurred at $T_S$. The overcooled clusters of Pnma phase are presumably present in $P2_1/m$ matrix at low temperatures due to the first order character of structural transformation.

The role played by these clusters is what distinguishes the physical properties of Sm doped from those of the Sm-Nd doped manganites. In $Ca_{1-x}Sm_xMnO_3$ series, no other singularities are clearly seen in the physical properties except those in the vicinity of structural phase transition (cf. Fig. 1 of Ref. [17] and Fig. 1 of present work). It seems, that in the more disordered manganites, namely, in $Ca_{1-x-y}Sm_xNd_yMnO_3$ samples the increase in the spread of the size of the doping elements leads to an enhancement of phase separation aspects in resistivity and magnetization measurements.

Assuming that ferromagnetic clusters and anti-ferromagnetic clusters of G-type are frozen in anti-ferromagnetic matrix of C-type the step-like upturn in magnetization and metal-insulator transition at $T_{MI}$ could be attributed to the ferromagnetic ordering in clusters of orthorhombic Pnma phase. The sharp increase in resistivity at this temperature is of special importance to get better insight into physical properties of this non-uniform state. The ferromagnetic transition in the clusters would not influence significantly the sample's resistivity without the redistribution of carrier density. The ferromagnetic ordering in these clusters is accompanied probably by the "sponging" of the conduction electrons from the neighboring anti-ferromagnetic matrix leading the system through percolation threshold. This process could be favoured by the gain in the kinetic energy of electrons. Note, that in the more uniform $Ca_{0.80}Sm_{0.20}MnO_3$ system, the transition into insulating state coincides with the structural phase transition temperature (cf. Fig. 2 of Ref. [17] and Fig. 6,7,8 of present work). The absence of anomalies in specific heat and thermal expansion indicates that the ferromagnetic ordering occurs only in minor parts of the sample, not in the matrix itself. These observations once again allow us to exclude the scenario of the transition into canted anti-ferromagnetic state in bulk. Otherwise, the appearance of ferromagnetic component in basically anti-ferromagnetic matrix would favour the decrease of resistivity. The analytical calculations[20] support the phase separation scenario to that of canted antiferromagnetism.

The size of these clusters in $Ca_{1-x-y}Sm_xNd_yMnO_3$, if any, appears to be an open question. The concept of electronic phase separation at nanoscale level goes to early suggestions of ferrons[21], where it was postulated that if two phases have opposite charge, the Coulomb forces break the macroscopic clusters into microscopic ones. While it is not obvious at all that the concept of ferrons is applicable to manganites discussed here , the experimental observations suggest the presence of large-scale inhomogeneities with percolative and fractal properties [7]. The submicrometer inhomogeneities could not survive, once the long-range Coulomb interaction is taken into account. The possible mechanism for the large-scale clusters formation was recently proposed [22]. The essence of this idea is that the neighbouring phases possess different patterns of spontaneous symmetry breaking, so that a transition between them is of first order.

The complex lanthanide doping of electron manganites undoubtedly increases the disorder in the system. Of course, it appears for substitution of Ca by any Ln, since these ions possess different charge. But, even the mismatch effect at doping by isovalent Sm and Nd will lead to additional disorder. In fact, the submicromete clusters were observed[7] in the compounds with complex lanthanides doping, $(La_{1-y}Pr_y)_{1-x}Ca_xMnO_3$. The Monte Carlo simulations [22] indicate that weak disorder in the vicinity of first order transition leads to cluster coexistence at a scale much larger than nanoscale one [6].

## VII. CONCLUSION

The experimental data presented here give clear evidence of the enhancement of various phase separation effects in electron manganites with an increased A- site disorder. The comparison with less disordered samples of $Ca_{1-x}Sm_xMnO_3$ series allows us to propose that the origin of these effects lies in the increased spread in sizes of trivalent lanthanide cations interpolated into perovskite structure. The signatures of phase separation appear to be more pronounced in the overdoped $Ca_{0.8}Sm_{0.16}Nd_{0.04}MnO_3$ and $Ca_{0.8}Sm_{0.18}Nd_{0.02}MnO_3$ samples as compared to the optimally doped $Ca_{0.85}Sm_{0.08}Nd_{0.07}MnO_3$ and $Ca_{0.85}Sm_{0.10}Nd_{0.05}MnO_3$ samples. The ground state of $Ca_{0.8}Sm_{0.16}Nd_{0.04}MnO_3$ is assumed to be that of metallic ferromagnetic clusters and insulating anti-ferromagnetic clusters of G-type embedded into insulating anti-ferromagnetic matrix of C-type. The metal-insulator transition at $T_{MI} \sim 112 \pm 3$ K occurring well below the structural phase transformation at $T_S \sim 158 \S 4$ K could be favoured by "sponging" of carriers by ferromagnetic clusters leaving the matrix insulating. The direct observation of metallic ferromagnetic inclusions on the insulating anti-ferromagnetic background could be of primary importance to check the "sponging" effect discussed.

The balance between competing phases in electron doped manganites seems to be rather subtle, since small changes in composition lead to large variations in their properties as demonstrated here. The enhancement of phase separation aspect in complex doped electron manganites should be taken into consideration at fine tuning and optimization of physical properties of colossal magnetoresistacne materials. And finally, we would like to point out that the data presented here would be of considerable significance in the light of the recent work by Dagotto [23] wherein the author discusses the relevance of clustered states to the physics of CMR manganites.

This work was supported by NWO grant 008-012-047, RFB grant 02-02-16636, RFB-BRFBR 02-02-81002 and RFBR 03-02-16108.

References

[1] *Colossal Magnetoresistance, Charge Ordering and Related Properties of Manganese Oxides*, ed C N R Rao and B Raveau (World Scientific, Singapore,1998).

[2] *Colossal Magnetoresistive Oxides*, ed Y Tokura (Gordon and Breach, New York, 2000).

[3] Rodriguez-Martinez L M and Attfield J P 1996 *Phys. Rev*. B **54** R15622

[4] Moreo A, Yunoki S and Dagotto E 1999 *Science* **283** 2034

[5] Salamon M B and Jaime M 2001 *Rev. Mod. Phys*.**73** 583

[6] Dagotto E 2002 *Nanoscale Phase Separation and Colossal Magnetoresistance.* (Springer-Verlag, New York).

[7] Uehara M, Mori S, Chen C H, and Cheong S W 1999 *Nature* **399** 560

[8] Maignan A, Martin C, Damay F and Raveau B 1998 *Chem. Mater*. **10** 950

[9] Raveau B, Maignan A, Martin C, and Hervieu M 1998 *Chem. Mater*. **10** 2641

[10] Raveau B, Maignan A, Martin C and Hervieu M 1999 *J.Supercond*. **12** 247

[11] Maignan A, Martin C, Damay F, Raveau B and Hejmanek J 1998 *Phys. Rev*. B **58** 2758

[12] Neumeier J J and Cohn J L 2000 *Phys. Rev*. B **61** 14319

[13] Mahendiran R, Maignan A, Martin C, Hervieu M, and Raveau B 2000 *Phys. Rev*. B **62** 11644

[14] Martin C, Maignan A, Hervieu M, Raveau B, Jirak Z, Kurbakov A, Trounov V, Andre G, Bouree F 1999 *J. Magn. Magn. Mater*. **205** 184

[15] Respaud M, Broto J M, Rakoto H, Vanacken J, Wagner P, Martin C, Maignan A and Raveau B 2001 *Phys. Rev*. B **63** 14426

[16] Lagutin A S, Vanacken J, Semeno, Bruynseraede Y and R. Suryanarayanan R 2003 *Sol. St. Comm*. **125** 7

[17] Martin C, Maignan A, Damay F, Hervieu M and Raveau B 1997 J. Solid State Chem. **134** 198

[18] Algarabel P A, De Teresa J M , García-Landa B, Morellon L, and Ibarra M R, Ritter C, Mahendiran R, Maignan A, Hervieu M, Martin C, Raveau B, Kurbakov A and Trounov V 2002 *Phys. Rev*. B **65** 104437

[19] Filippov D A, Levitin R Z, Vasil'ev A N, Voloshok T N, Kageyama H and Suryanarayanan R 2002 *Phys. Rev.* B **65** R100404

[20] Kagan M Yu, Khomskii D I and Mostovoy M V 1999 *Eur. J. Phys..* B **12** 217

[21] Nagaev E L 1967 *JETP Lett*. **6** 18

[22] Moreo A, Mayr M, Feiguin A, Yunoki S, and Dagotto E 2000 *Phys. Rev. Lett*. **84** 5568

[23] Dagotto E *cond-mat*/0302550.

Figure Captions

Figure. 1: Magnetization as a function of temperature of $Ca_{1-x-y}Sm_xNd_yMnO_3$ of four different compositions in H =0.1 T.

Figure. 2: Magnetization as a function of temperature of $Ca_{0.8}Sm_{0.16}Nd_{0.04}MnO_3$ in zero field-cooled (ZFC) and field- cooled (FC) regimes. Inset: The temperature dependence of low field ac magnetic susceptibility.

Figure. 3: Magnetization as a function of field of $Ca_{0.8}Sm_{0.16}Nd_{0.04}MnO_3$ at $T > T_S$ (panel a) and at $T < T_S$ (panel b). Inset: The magnetic hysteresis loop at T = 5 K.

Figure. 4: Temperature dependence of susceptibility and inverse paramagnetic susceptibility of $Ca_{0.8}Sm_{0.16}Nd_{0.04}MnO_3$. Inset: The Temperature dependence of spontaneous magnetization.

Figure. 5: The field dependence of magnetization of $Ca_{0.8}Sm_{0.16}Nd_{0.04}MnO_3$ taken in pulsed magnetic field.

Figure. 6: Resistivity of $Ca_{0.8}Sm_{0.16}Nd_{0.04}MnO_3$ as a function of temperature

Figure. 7: Thermal expansion of $Ca_{0.8}Sm_{0.16}Nd_{0.04}MnO_3$ as a function of temperature

Figure 8. Specific heat of $Ca_{0.8}Sm_{0.16}Nd_{0.04}MnO_3$ as a function of temperature

Figure 1

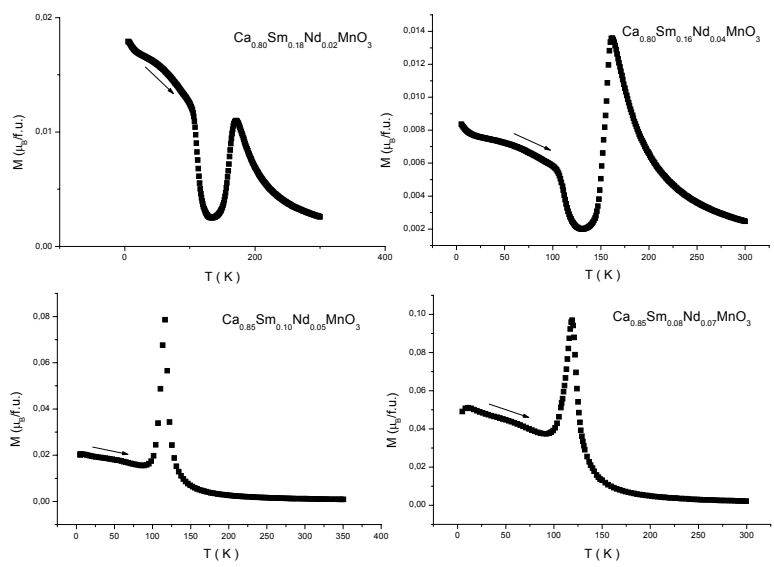

Figure 2

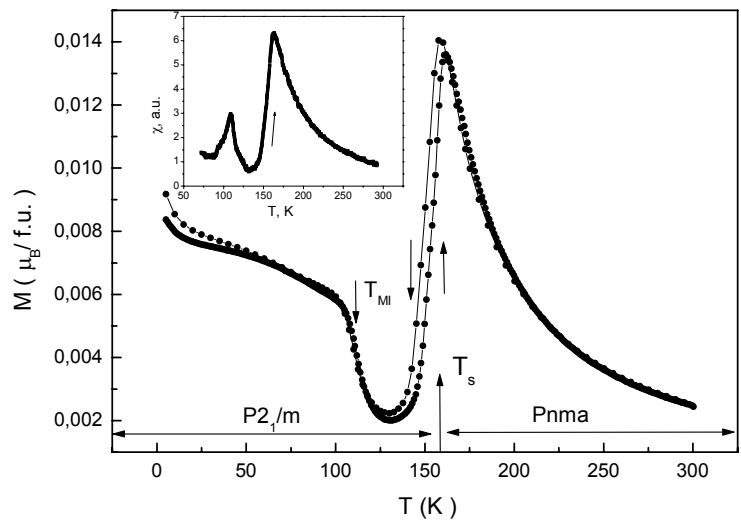

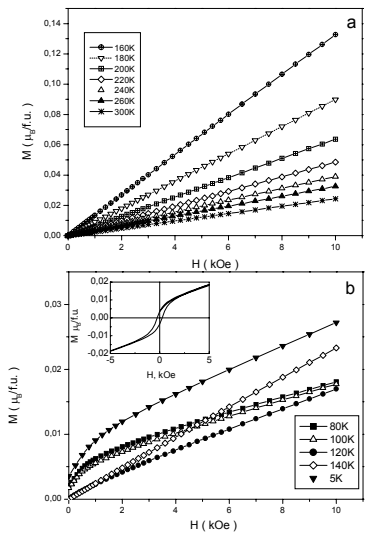

Figure 4

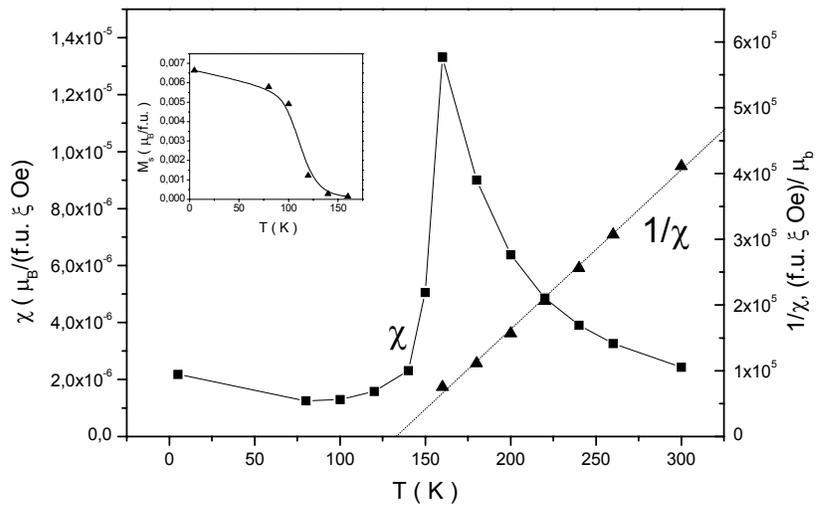

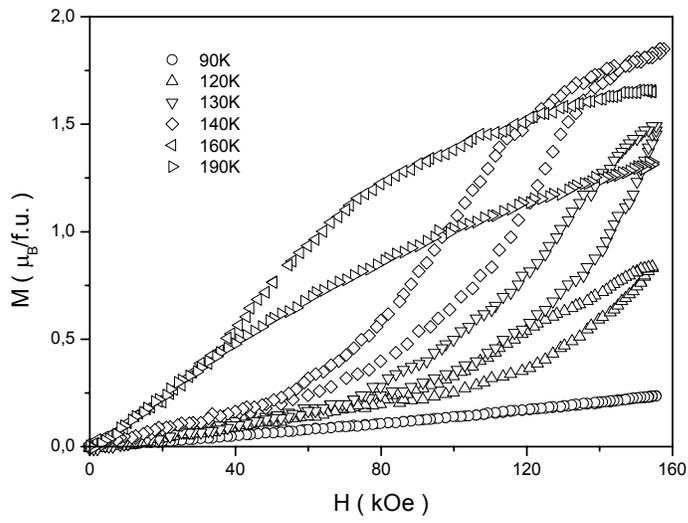

Figure 6

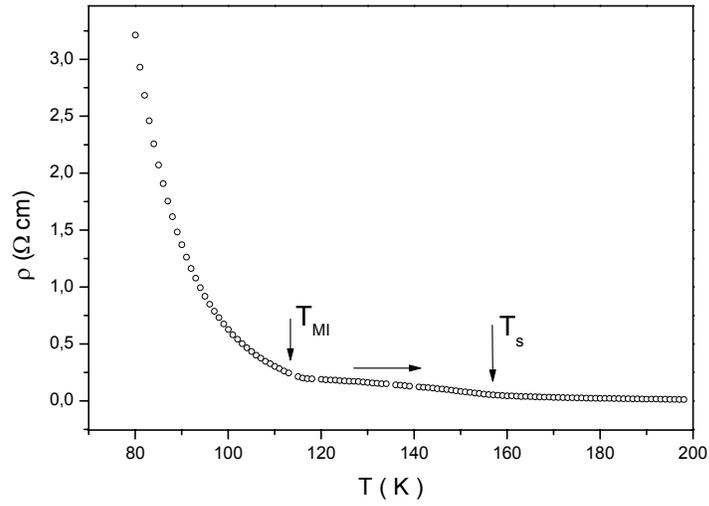

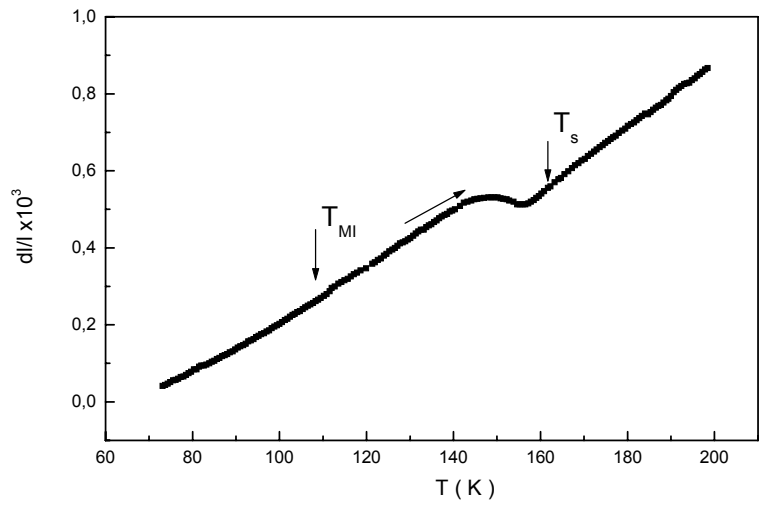

Figure 8

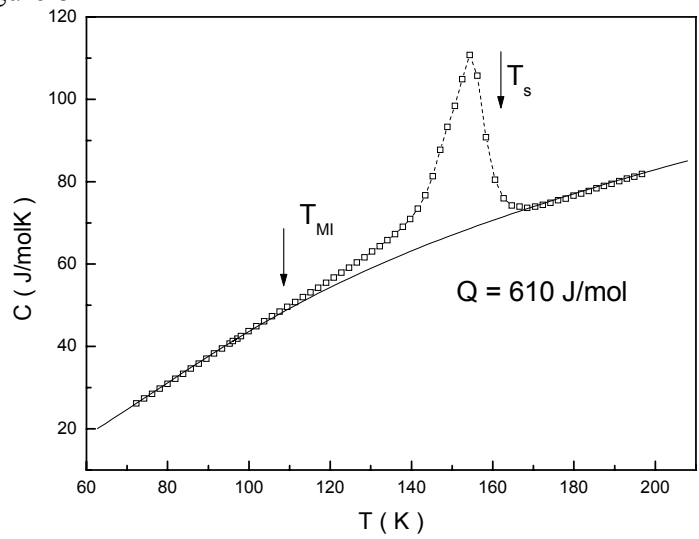